\documentclass[
]{ceurart}

\usepackage{listings}
\begin{document}

\copyrightyear{2026}
\copyrightclause{Copyright for this paper by its authors.
  Use permitted under Creative Commons License Attribution 4.0
  International (CC BY 4.0).}

\conference{Italian Workshop on Artificial Intelligence for Human-Machine Interaction (AIxHMI 2026), October 6-9, 2026, Perugia, Italy}

\title{Designing Human-mediated AI Guidance: Ready Together for Personalized Family Emergency Preparedness}

\author[1]{Nini Kurashvili}[%
email=n.kurashvili1@campus.unimib.it,
]
\cormark[1]
\fnmark[1]
\address[1]{University of Milano-Bicocca, Milan, Italy}

\author[1]{Yana Ivanchenko}[%
email=y.ivanchenko@campus.unimib.it,
]
\fnmark[1]

\author[1]{Greta Schiavo}[
email=gretaschiavo321@gmail.com]
\fnmark[1]

\author[1]{Cansu Koyuturk}[%
orcid=0009-0005-7562-7400,
email=cansu.koyuturk@unimib.it,
]

\author[1]{Dimitri Ognibene}[%
orcid=0000-0002-9454-680X,
email=dimitri.ognibene@unimib.it,
]

\cortext[1]{Corresponding author.}
\fntext[1]{These authors contributed equally.}

\begin{abstract}
Artificial intelligence (AI) systems are increasingly used across domains to provide personalized information, recommendations, and decision support. However, in some contexts, AI-generated information may not be suitable for direct delivery to the final recipient. Instead, it may need to be interpreted, adapted, and communicated by a human who understands the recipient’s needs, emotional state, and situational context. Human–AI interaction research has given less attention to situations in which a more knowledgeable human acts as an intermediary between an AI system and a less experienced or less informed recipient. We introduce \textit{the human-mediated AI guidance framework} and explore it through Ready Together, an AI-supported family emergency preparedness system in which parents mediate AI-generated content for their children. The system is designed to provide personalized guidance and support parents in making emergency preparedness more interactive and understandable through guided activities and family-centered learning. The system design was informed by a qualitative, design-oriented research process involving semi-structured interviews and co-design activities. Findings identified challenges in family emergency preparedness, including difficulty discussing emergencies with children, uncertainty about providing appropriate explanations, and a preference for interactive learning activities. These findings informed the design of an interactive prototype, subsequently evaluated through a pilot study and a heuristic evaluation. Participants responded positively to the personalized recommendations and practical activities. Preliminary findings suggest that human-mediated AI guidance may support context-sensitive family preparedness while preserving parents' responsibility for interpreting, adapting, and communicating AI-generated information.

\end{abstract}

\begin{keywords}
  Large language models \sep
  Emergency preparedness \sep
  Child-LLM interaction \sep
  Parental mediation \sep
  Human–AI collaboration \sep
  Personalized learning
\end{keywords}

\maketitle

\section{Introduction}
Emergencies such as fires, floods, earthquakes, and other crisis events can occur unexpectedly and can affect both physical safety and psychological well-being \cite{roudini2017}. Such events may cause stress and uncertainty, while effective responses require people to assess threats and make appropriate protective-action decisions \cite{roudini2017,lindell2012}. In these situations, families need clear and practical guidance to assess risks, plan appropriate actions, and maintain safety \cite{lindell2012}.
Studies on disaster preparedness suggest that advance preparation can support more informed protective-action decisions and improve readiness to respond to hazards \cite{paton2003,lindell2012}. Preparedness may include planning protective actions, identifying responsibilities, and ensuring access to necessary resources. However, people often delay such actions because preparedness requires present effort for uncertain future benefits \cite{linnemayr2016}. 
Accordingly, preparedness can strengthen families’ capacity to anticipate hazards and take appropriate protective actions \cite{paton2003,lindell2012}. While emergency preparedness is important, many families remain underprepared for potential crises \cite{paton2003}. Parents may find it difficult to discuss emergency-related topics with their children because communicating threatening information requires balancing preparedness with children’s emotional needs \cite{midtbust2018}. In addition, children’s learning needs should be considered, as educational theories emphasize active exploration, social interaction, and practical experience as important processes in learning \cite{kolb1984,piaget1952,vygotsky1978}. Therefore, there is a need for approaches that support parents in preparing their children for emergencies in an appropriate, clear, and supportive way.

Recent advances in large language models (LLMs) create new opportunities for personalized education, guidance, and decision-support systems. AI systems can adapt information and generate recommendations based on user needs; however, AI-generated guidance alone may not be sufficient in sensitive contexts such as family emergency preparedness. Prior work on human–AI interaction emphasizes the importance of designing AI systems that support user control, interpretation, correction, and appropriate reliance rather than treating AI output as automatically final \cite{amershi2019}. 
However, human--AI interaction is commonly designed around a direct interaction in which the person operating the AI is also the intended recipient of its output. This assumption becomes problematic when AI-generated information is intended for another person who may have different informational, contextual, or emotional needs and may not be able to express the input or evaluate the output independently.
In such settings, effective AI guidance depends partly on communicating relevant information about the intended recipient to the system.
However, users, especially those with limited experience using LLMs, may struggle to give clear, sufficiently detailed, and well-structured inputs \cite{koyuturk2025understanding,zamfirescu2023johnny}. Effective prompting often requires users to specify goals, constraints, assumptions, and contextual information, suggesting that systems should support users in framing the situation rather than assuming they already know what information to provide \cite{martinenghi2026genie}. This is particularly important because LLM responses can be strongly influenced by the user’s initial framing, including incomplete or flawed inputs \cite{koyuturk2026hidden}.
Therefore, AI systems should not only generate guidance, but also support the human mediator in framing the situation by asking relevant questions and helping structure the necessary context. This idea is consistent with research on parental mediation, which conceptualizes parents as active interpreters and guides of children’s media and technology experiences \cite{clark2011}, as well as recent work on family use of generative AI, which shows that parents mediate children’s AI use through supervision, guidance, co-use, and rule-setting \cite{zhang2025families}.

Taken together, these considerations suggest the need to move beyond a direct human--AI interaction model toward a \textbf{human-mediated AI guidance model}. In this model, a more knowledgeable or context-aware human acts between an AI system and a final recipient. The mediator performs a bidirectional role: representing the recipient to the AI by providing information about their needs, abilities, emotional state, and situation, and subsequently evaluating, selecting, interpreting, and adapting the generated content before communicating it to the recipient. The mediator is therefore not simply a messenger, but functions as a context provider, reviewer, emotional adapter, and communicator while retaining responsibility for deciding how and whether AI-generated content should be used.

We investigate this model through \textit{Ready Together}, an AI-supported family emergency-preparedness system in which parents mediate AI-generated guidance for their children. Parents provide relevant information about the child and family, including the child's age, level of understanding, emotional needs, previous experiences, and the emergency scenario for which the family wants to prepare. Based on this information, the AI generates personalized explanations, practical recommendations, family activities, role-playing exercises, and checklists that parents can review and adapt before using them with their children. The system is thus intended to support, rather than replace, parental judgment and communication.

Human-mediated AI guidance, as instantiated here, is one case of a more general phenomenon we term \emph{human-mediated AI interaction}: settings in which AI-generated information is intended for a person other than the operator. This configuration raises questions that are less visible when human--AI interaction is modeled primarily as a dyadic exchange:

\begin{itemize}
\item[(a)] Can users translate AI-generated information when it requires contextual, relational, or emotional adaptation before reaching the intended recipient?
\item[(b)] How can a human convey contextual, social, relational, and emotional information about the intended recipient to an AI system?
\item[(c)] How can the AI system support this mediation process, both by eliciting the context it needs and by structuring its output so that the mediator can adapt and present it appropriately?
\item[(d)] How does interacting with the AI affect the mediator's ongoing interaction with the intended recipient and the broader social context in which the guidance is delivered?
\end{itemize}

The present study provides initial observations concerning (a) and (b) within a single domain and offers design observations relevant to (c). Question (d) emerges from our design analysis but is not evaluated in the present preliminary study.

We conducted a qualitative, design-oriented study involving semi-structured interviews and co-design activities with six parents. The findings informed the design of the Ready Together prototype and an AI guidance component implemented through a Custom GPT. The system concept was subsequently examined through a pilot study and a heuristic evaluation of the interface. The primary contribution of this work is the \textbf{human-mediated AI guidance framework}, supported by preliminary observations about parents' experiences of mediating AI-generated guidance and resulting design implications for mediator--AI--recipient systems.

\section{Background}

\subsection{Psychological Barriers to Emergency Preparedness}
Emergency preparedness is an important prerequisite for family safety and resilience. However, many families can postpone or do not start preparing at all. Research shows that this is not solely due to a lack of information or resources. Inadequate risk perception, uncertainty about available protective actions, decision-making delays, and situational impediments can also prevent families from preparing in a timely manner \cite{lindell2012}.

Cognitive biases and emotional factors also influence how families assess risk and respond to it. 
One relevant factor is optimism bias, or the tendency to believe that negative events are more likely to affect other people than oneself \cite{weinstein1980}. When families perceive an emergency as unlikely to affect them personally, they may underestimate the value of advance preparation and delay preventive action \cite{najafi2017}. A related tendency is normalcy bias, in which people assume that ordinary conditions will continue despite warning signs or potential threats \cite{mileti1992}. This can lead individuals to minimize risk or postpone protective behavior until a situation appears immediate \cite{mileti1992,nationalresearchcouncil2006}.
Preparedness can also be affected by temporal discounting. People often assign greater importance to immediate tasks and benefits than to actions intended to reduce uncertain future risks \cite{frederick2002}. Although emergency planning may be viewed as valuable, it can be displaced by everyday responsibilities that appear more urgent. As a result, preparedness may remain an acknowledged but incomplete family goal.
Emotional avoidance creates an additional challenge, particularly when parents consider discussing emergencies with children. Conversations about fires, disasters, evacuations, or other threatening situations may produce fear, anxiety, or discomfort \cite{lazarus1984}. Parents may therefore avoid these conversations because they are concerned about frightening their children or are uncertain about how to explain risks calmly. However, avoiding the topic may leave children without age-appropriate knowledge about safety and emergency behavior. Research on disaster communication emphasizes the importance of child-centered education and parental guidance in helping children understand risks and prepare for emergencies \cite{midtbust2018}.

Together, these barriers suggest that an effective family preparedness intervention should do more than provide general safety information. It should make preparedness personally relevant, reduce the emotional and practical burden of beginning the process, and divide preparation into manageable actions that can be incorporated into family life.

\subsection{Learning Theories Supporting Family Preparedness}
In addition to psychological barriers, it is important to consider how learning occurs in the family environment, especially for children. According to educational theories, learning by doing is more effective when it is based on active participation, reinforced by social interaction, and linked to real or simulated experiences, rather than just passively receiving information.
Constructivism \cite{piaget1952} states that children actively construct knowledge through interaction with their environment, not just by passively receiving information. Learning is most effective when children are involved in exploratory activities, decision-making, and engaging in activities that allow them to internalize knowledge in real-world situations. From this perspective, preparedness education should involve opportunities to explore situations, make decisions, and practice appropriate responses. Rather than only being told what to do during an emergency, children may benefit from participating in simulations, selecting safe actions, assembling preparedness materials, or discussing what could happen in a familiar setting.
Social constructivism emphasizes that children learn through interaction with more knowledgeable others \cite{vygotsky1978}. In the family context, parents can play this role by supporting the child’s understanding through explanation, guidance, and contextualization. Accordingly, in the application, parents act as mediators, adapting the AI-generated content to the child’s age and emotional needs. In this way, emergency preparedness becomes a shared family learning process rather than information that is simply delivered to the child.
Experiential learning similarly highlights the importance of experience, action, and reflection in developing knowledge \cite{kolb1984}. In the context of emergency preparedness, role-playing exercises, family drills, practical tasks, and guided discussions can make safety procedures more concrete.

Together, these theories motivate the use of active, social, and experiential activities in Ready Together, with AI providing structured activities and parents contextualizing and adapting them for their children.

\subsection{Human-Mediated AI Guidance}

AI systems are increasingly used to provide personalized information, recommendations, and decision-support across education, family life, health, and safety-related contexts. Much of the existing human–AI interaction literature focuses on direct interaction between a human user and an AI system, where the same person who receives the AI output is also expected to interpret, evaluate, and act on it. For example, guidelines for human–AI interaction emphasize the importance of user control, feedback, correction, and appropriate trust in AI systems \cite{amershi2019}. Similarly, human-centered AI research argues that AI should be designed to augment human abilities while preserving human control, responsibility, and trustworthiness \cite{shneiderman2020}. Work on AI-assisted decision-making also highlights the importance of complementarity between human and AI capabilities, accurate human mental models of AI, and appropriate reliance on AI recommendations \cite{steyvers2024,schemmer2023}.

However, these approaches usually focus on the dialogic interaction. Less attention has been given to situations in which the AI-generated information is intended for another human recipient, such as a child, student, novice, client, or other person who may not be able to formulate the input and evaluate the AI output independently. This idea is strongly connected to parental mediation theory, which explains how parents guide, regulate, and interpret children’s experiences with media and digital technologies \cite{clark2011}. Recent research extends this perspective to generative AI. Families may use and mediate generative AI in different ways, including restrictive, instructive, co-use, hands-on, and independent approaches \cite{zhang2025families}. Similarly, \cite{xie2026} examines parents’ perspectives on children’s use of AI for self-directed learning and shows that parents often monitor children’s AI use, reinterpret AI outputs, guide children’s prompting practices, and gradually adjust the level of independence children are allowed to have. These studies suggest that AI mediation in families is not only about controlling access, but also about helping children understand, evaluate, and use AI responsibly. 

HCI work on parent–AI collaboration also demonstrates the importance of keeping parents involved in AI-supported child learning. For example, StoryBuddy is a human–AI collaborative storytelling system designed to support parent-child interactive reading \cite{storybuddy2022}. It allows parents to configure question types, select or edit AI-generated questions, and decide how much AI support to use. This shows that parents bring contextual knowledge that the AI lacks and that children cannot readily share, including knowledge of the child’s abilities, interests, emotions, and learning goals. Similarly, a previous study \cite{mei2026adapting} has shown that parent–AI collaboration in real-time conversations with children is dynamic and context-dependent. Parents adjust how AI participates based on emotional intensity, conversation stage, parental goals, and the child’s needs. 

While prior work has examined parental mediation of digital media, family use of generative AI, and human-AI decision-making, \textit{Ready Together} is designed based on a framing of human-mediated AI guidance. In this model, the more knowledgeable other (parent) is not only a user or supervisor of AI, but a bidirectional mediator who first frames the recipient’s (child’s) needs, abilities, and family situation into a meaningful context for the AI, and then interprets, filters, emotionally adapts, and communicates AI-generated preparedness guidance to the child. Therefore, the mediator is not only a messenger who passes AI content forward, but is also a context provider, evaluator, ethical gatekeeper, and communicator.

\section{Research and Design Methods}
The project followed a qualitative, design-oriented research and development process, complemented by a pilot evaluation with descriptive quantitative results. First, a needfinding and user research stage was conducted to investigate family emergency preparedness practices, parent-child communication patterns, and parents’ perceptions of their children’s learning needs. Second, the findings from this stage informed the design and implementation of Ready Together as a human-mediated AI guidance system. Finally, a formative pilot evaluation was conducted to explore participants’ perceptions of the system concept and of the usefulness, clarity, and relevance of the AI-generated guidance. The evaluation was exploratory and was not intended to establish system effectiveness or produce generalizable findings.

\subsection{Needfinding and User Research}

The needfinding and user research aimed to understand how families perceive and value emergency preparedness, what challenges parents face when discussing emergency situations with their children, and how AI could support parents in creating child-appropriate preparedness guidance.
The needfinding process included two main methods: semi-structured interviews and co-design activities, which were conducted with six parents recruited through convenience sampling: five mothers and one father. Participants came from diverse cultural backgrounds, including Ukraine, Italy, Spain, Sweden, and Russia, and were aged between 28 and 55 years.

\textbf{Semi-Structured Interviews}:
They were used as the main needfinding method. The interview protocol was informed by previous research on emergency preparedness, psychological barriers, family resilience, parent-child communication, and children’s learning needs \cite{lazarus1984,lindell2012,kolb1984,vygotsky1978}. 
The interviews aimed to understand parents’ experiences, perceptions, and current preparedness practices. Participants were asked how they perceived emergency situations, whether they discussed such situations with their children, how prepared they felt, and what challenges they faced when sharing safety-related information with their children. The semi-structured format allowed participants to share personal experiences, concerns, and examples from everyday life, while also giving the researcher flexibility to ask follow-up questions.

\textbf{Co-Design Activities}:
After the interviews, participants were introduced to the initial Ready Together concept. They were asked to evaluate the proposed solution, provide feedback, identify potential problems, discuss useful features, and suggest activities that could be engaging for both children and parents. This co-design process allowed participants to contribute to the development of the system and helped ensure that the design reflected real family needs.

\subsection{Findings}

\subsubsection{Understanding Emergency Preparedness}
Participants identified a variety of emergencies, including: flood, fire, storm, earthquake, sudden illness, and emergency evacuations. Ukrainian families also reported war-related threats and air-raid sirens, suggesting that people's experiences and environments influence what they perceive as an emergency. 

\subsubsection{Challenges Experienced by Parents and Children}
Parents wanted their children to be well-prepared for emergencies. However, they often found it difficult to discuss these topics with their children because they were concerned that such conversations might frighten them. Some parents also reported that they could not find a clear and simple explanation, understandable for a child. Also, a lack of resources reduced their confidence level. 
They also perceived interactive, visual, and practical activities as more suitable for their children than text-based information alone.

\subsubsection{Behavioural Patterns Influencing Preparedness}
Participants’ descriptions were consistent with preparedness barriers discussed in prior literature \cite{lindell2012,slovic2010,weinstein1980,frederick2002}. Some described preparedness as less urgent than everyday responsibilities, expected normal conditions to continue, or viewed preparedness education as primarily the school’s responsibility. These perceptions contributed to delaying family preparedness activities.

\subsection{Ready Together: System Design and Interactive Prototype}

\begin{figure}[htbp]
  \centering
  \includegraphics[width=\linewidth]{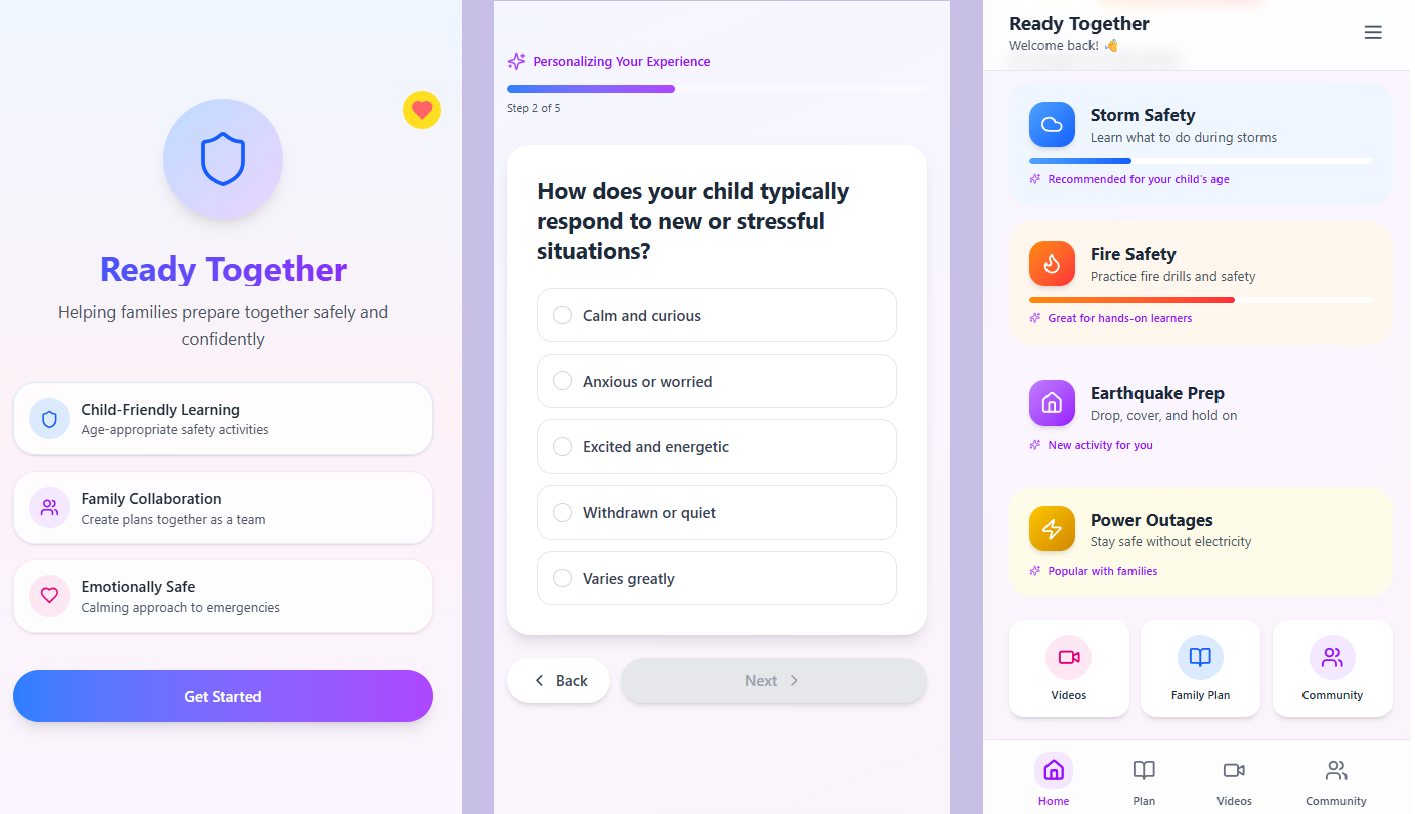}
  \caption{Example screens from the prototype, including the welcome screen, personalization process, and emergency-preparedness activity dashboard.}
  \label{fig:ready-together}
\end{figure}

The Ready Together interface was developed as an interactive Figma prototype representing the intended application structure and user flow (Figure\ref{fig:ready-together}); it was not implemented as a fully integrated mobile application. The prototype includes five conceptual components: the parent interface, AI guidance module, family activity module, family preparedness plan, and experience-sharing form.

The first component is the parent interface. This is where parents provide contextual information about the child to create a family profile, such as age, emotional needs, level of understanding, the family context, such as previous experience, and the family’s current preparedness level and the emergency scenario they want to prepare for. The system supports this process through structured prompts and input fields, while also allowing parents to add extra information manually. 

The second component is the AI guidance module. In this module, AI uses the information provided by the parent to generate personalized preparedness content, including explanations, recommendations, activities, role-playing exercises, and checklists. 
In the pilot study, the AI guidance module is implemented as a Custom GPT on the ChatGPT platform and operates separately from the Figma interface. The intended integrated version of Ready Together would connect the parent-facing interface to an LLM-based guidance service. Information entered through the family profile and emergency-preparation flow would be structured as contextual input to the model, together with system-level instructions defining the assistant's role and output structure.

The third component is the family activity module, which turns preparedness guidance into interactive learning activities, such as games, discussions, simulations, and practical exercises. These activities are designed to help children learn safety-related information in a more engaging and hands-on way.

The fourth component is the family preparedness plan. This part helps families organize important emergency information, such as emergency contacts, safe places, family roles, supplies, and step-by-step actions for different emergency situations.

The fifth component is the experience-sharing form. This form allows parents to describe their own preparedness experiences in a structured way, including the emergency scenario, actions taken, challenges faced, and lessons learned. These shared experiences can help other parents reflect on real family situations and improve their own preparedness.

\subsection{Formative Pilot Evaluation}

Ready Together was examined through a formative pilot evaluation using a demonstration prototype. The purpose of the pilot was to explore how participants perceived the system concept and the AI-generated preparedness guidance, rather than to evaluate the effectiveness of the system or its impact on preparedness outcomes. The Figma prototype represented the intended interface and user flow, while the AI guidance component was implemented separately as a Custom GPT accessed through the ChatGPT platform. It was configured with a system prompt that simulated the intended in-app behavior by collecting information from parents to provide family, child, and emergency context to generate personalized preparedness guidance in a multi-turn interaction.

The pilot study involved the six parents who had previously participated in the needfinding and co-design stages. During the study, participants interacted with the Ready Together AI assistant and were asked to describe an emergency situation they wanted to prepare their family for, such as a fire, storm, power outage, evacuation, or a child getting lost in a public place. Based on the information provided by the parent, the assistant generated personalized preparedness guidance, including explanations intended for children, practical recommendations, family activities, role-playing exercises, and simple emergency checklists. Later, participants reviewed the content and evaluated it by answering five-point rating-scale items and open-ended questions.

\subsubsection{Quantitative Results}
The evaluation focused on parents’ perceptions of the guidance’s clarity, usefulness, relevance, potential to support child preparedness, and their intention to use the system in the future. Given the exploratory nature of the study and the small sample size, the ratings are reported only as descriptive summaries of participants’ perceptions.

Overall, they reported positive perceptions of the AI-generated recommendations (Table\ref{tab:pilot-results}). The recommendations were clear and easy to follow, suggesting that parents perceived the guidance as understandable and practical. Recommendations were also rated positively for relevance to participants’ family situations. These ratings represent preliminary descriptive feedback rather than evidence of system effectiveness.

\begin{table}[t]
\caption{Participants’ mean ratings of the AI-generated preparedness guidance}
\label{tab:pilot-results}
\centering
\begin{tabular}{@{}p{0.65\linewidth}r@{}}
\toprule
\textbf{Statement} & \textbf{Mean score} \\
\midrule
The recommendations matched my family’s situation & 4.3 \\
The recommendations were useful & 4.2 \\
The recommendations were clear and easy to follow & 4.5 \\
The recommendations would help me prepare my child for emergencies & 4.2 \\
I would use this application in the future & 4.0 \\
\bottomrule
\end{tabular}
\end{table}

\subsubsection{Qualitative Results}
Answers to the open-ended questions showed that participants positively evaluated the personalized nature of the AI-generated guidance. They noted that the recommendations were useful because they were connected to their family situation and the selected emergency scenario, rather than being general preparedness advice.
Participants also appreciated the practical format of the guidance. Family activities, role-playing exercises, and checklists helped transform emergency preparedness into concrete actions that could be used with children. This was important because many parents reported difficulties in explaining emergencies in a simple and calm way.
Another important point was related to the role of the parent as a mediator. Participants emphasized that AI-generated content should not be given directly to children. Instead, parents need to review and adapt the information according to the child’s age, emotional state, and level of understanding. This perception is consistent with the main design rationale of Ready Together, in which AI provides structured guidance while parents retain responsibility for reviewing and adapting the content.
However, some participants noted that the AI occasionally combined several types of information, such as an explanation, a checklist, and a family activity, within one long response. This made it more difficult for parents to quickly identify which parts were suitable to share with the child. In these cases, parents had to shorten the response, simplify some sentences, and select the most relevant recommendations before they considered the content suitable for use with their children. They also suggested that the system should include more emergency scenarios, more interactive materials, and additional support for parents.

\subsubsection{Design Improvements}
Based on the pilot study feedback, several design improvements were identified for Ready Together. First, the system should provide stronger support for parents in adapting AI-generated guidance. This may include examples of appropriate language, calming explanations, and short prompts that help parents explain emergencies in a clear and emotionally safe way. To address the difficulty, future versions should separate the guidance into clearly labeled sections, such as “Explanation for the Child,” “Advice for the Parent,” and “Family Activity.” The system should also allow parents to request a shorter version or adjust the level of detail before using the content with their children.
Second, the application should expand its emergency scenario library. Participants indicated that families may face different types of emergencies and a broader set of scenarios would make the system more relevant to different family contexts.
Also, Ready Together should offer more role-playing exercises, simulations, games, and preparedness tasks.
Finally, the system should improve usability by making instructions clearer, allowing parents to edit family profile information, and showing progress during longer activities. These improvements would make the system easier to use and would better support the parent's mediating role.

\subsubsection{Heuristic Evaluation}
A heuristic evaluation of the Ready Together prototype was conducted by one expert evaluator using Nielsen’s usability heuristics \cite{nielsen1994}. The evaluator assessed the prototype against Nielsen’s ten heuristics and assigned each heuristic a severity rating from 0 to 4, where 0 indicated no usability problem, and 4 indicated a severe usability problem.
The evaluator identified positive aspects of the prototype, including its family-centered workflow and simple visual design, alongside issues involving progress visibility, profile editing, accidental exits, and help documentation. Suggested improvements included progress indicators, profile editing, confirmation messages, quick access to scenarios, and additional guidance for parents.

\begin{table}[t]
\caption{Heuristic evaluation results of the Ready Together prototype}
\label{tab:heuristic-evaluation}
\centering
\begin{tabular}{@{}p{0.52\linewidth}r@{}}
\toprule
\textbf{Heuristic} & \textbf{Severity rating} \\
\midrule
Visibility of system status & 1 \\
Match between system and the real world & 1 \\
User control and freedom & 2 \\
Consistency and standards & 1 \\
Error prevention & 2 \\
Recognition rather than recall & 2 \\
Flexibility and efficiency of use & 2 \\
Aesthetic and minimalist design & 0 \\
Help users recognize, diagnose, and recover from errors & 1 \\
Help and documentation & 2 \\
\bottomrule
\end{tabular}
\end{table}

\section{Discussion}
This study introduced \textit{the human-mediated AI guidance framework} through Ready Together, an AI-supported system designed to help parents create emergency-preparedness guidance for children, and examined how parents perceived their role as mediators of AI-generated content. 
The formative findings suggest that participating parents perceived AI-generated emergency-preparedness guidance as relevant, understandable, and potentially useful for supporting conversations and activities with their children. Participants valued recommendations that reflected their family context and selected emergency scenario, as well as practical formats such as activities, role-playing exercises, and checklists. These results align with social constructivism, in which a more knowledgeable person supports and contextualizes learning \cite{vygotsky1978} and experiential learning, which emphasizes active practice rather than passive information delivery \cite{kolb1984}.The Ready Together prototype represents an interaction model in which parents provide contextual information about the child and family, AI-generated preparedness material is produced based on this context, and parents are expected to review and adapt the content before using it with their children. However, because the pilot assessed parents’ perceptions rather than preparedness outcomes, the findings should not be interpreted as evidence that Ready Together improves family preparedness or children’s understanding.

Beyond the Ready Together prototype, the conceptual contribution is a shift from direct human–AI interaction toward a three-actor model involving the AI, a human mediator, and an intended recipient. First, the mediator represents the intended recipient to the AI by providing contextual information that the recipient may be unable to provide independently, such as their abilities, needs, emotional condition, goals, and circumstances. Second, the mediator evaluates and transforms the AI output before communicating it to that recipient. The mediator therefore acts as a context provider, reviewer, interpreter, emotional adapter, and accountable decision-maker. 

This framework may also be relevant to domains in which the final recipient has less expertise, limited decision-making capacity, reduced access to relevant context, or difficulty evaluating AI-generated information. For example, in education, teachers may provide information about students’ prior knowledge, learning difficulties, language abilities, and classroom goals, and then adapt AI-generated explanations or exercises before presenting them to students. In workplace and customer-support settings, employees can provide organizational and client context to an AI system and then adapt the generated response before communicating with a customer or colleague. Nevertheless, the employee remains important for identifying unusual situations, correcting inappropriate suggestions, and adapting communication to the customer’s needs.
Across these domains, the framework suggests several common design requirements. Systems should help mediators construct relevant context, separate information for the mediator from content intended for the recipient, disclose uncertainty, and support escalation to qualified humans. Responsibility should remain visible rather than being obscured by fluent AI output. The framework is intended for contexts where AI supports the mediator while preserving human judgment, relational knowledge, and accountability.

\section{Limitations and Conclusion}
This preliminary study is an initial exploration of the human-mediated AI guidance framework using the “Ready Together” example, and therefore has several limitations.
First, the study had a limited number of participants, suggesting that the results may not be generalizable to all families. Also, the pilot participants were the same parents who had participated in the needfinding and co-design stages. Their familiarity with the concept may have influenced their evaluations, and the study therefore should be interpreted as a formative evaluation rather than an independent validation of the system.

Second, the study focused primarily on parents’ perceptions. Children did not participate directly, and the study did not measure children’s understanding, anxiety, knowledge retention, or ability to perform emergency actions. Future studies should include children through age-appropriate and ethically designed methods.
The study also examined the framework within a single domain and relied on an early prototype and a pilot evaluation. Future research should investigate how mediation roles differ across domains such as education, healthcare, and the workplace, how much effort mediation requires from users, and which interface mechanisms best support accurate, ethical, and context-sensitive adaptation. 

Overall, the formative findings suggest that participating parents perceived human-mediated AI guidance as a potentially useful approach for creating preparedness content that they could review and adapt for their children. The study also highlights the importance of preserving human judgment and responsibility when AI-generated guidance is intended for a recipient who may not be able to evaluate it independently, supporting the proposed human-mediated AI guidance framework as a way of conceptualizing this intermediary role.

\begin{acknowledgments}
This work was supported by the Italian Ministry of University and Research under Grant No. 2023-NAZ-0206 and PsyFuture – Dipartimento di Eccellenza 2023-2027.
\end{acknowledgments}

\section*{Declaration on Generative AI}
During the preparation of this work, the author(s) used OpenAI’s ChatGPT for grammar correction and to refine phrasing and improve the readability of the text. After utilizing this tool, the author(s) thoroughly reviewed and edited the content as necessary and took full responsibility for the content of the published article.

  

\bibliography{references}


\end{document}